\documentclass[10pt,conference]{IEEEtran}
\IEEEoverridecommandlockouts
\usepackage[T1]{fontenc}
\usepackage{cite}
\usepackage{amsmath,amssymb}
\usepackage{graphicx}
\usepackage{booktabs}
\usepackage{multirow}
\usepackage{array}
\usepackage{enumitem}
\usepackage{csquotes}
\usepackage{tcolorbox}
\usepackage[numbers]{natbib}
\definecolor{quotepurple}{RGB}{94,52,140}
\newcommand{\pq}[1]{{\color{quotepurple}\enquote{\textit{#1}}}}
\tcbset{takeaway/.style={colback=black!4,colframe=black!55,boxrule=0.4pt,arc=1.5pt,left=4pt,right=4pt,top=3pt,bottom=3pt,before skip=6pt,after skip=6pt}}
\usepackage{xcolor}
\usepackage{colortbl}
\definecolor{zfaster}{HTML}{A7C2E8}\definecolor{zfurther}{HTML}{C4D9F3}\definecolor{znew}{HTML}{CFE6F6}\definecolor{phBefore}{HTML}{DCEFFA}\definecolor{phWhile}{HTML}{FBE1D9}\definecolor{phWrong}{HTML}{FFF1CC}\definecolor{phAfter}{HTML}{E7F1D2}\definecolor{zfurtherA}{HTML}{E8F1FC}\definecolor{zfurtherB}{HTML}{D7E6F8}
\usepackage[hidelinks]{hyperref}
\usepackage{microtype}
\def\BibTeX{{\rm B\kern-.05em{\sc i\kern-.025em b}\kern-.08em T\kern-.1667em\lower.7ex\hbox{E}\kern-.125emX}}

\begin{document}
\bstctlcite{IEEEexample:BSTcontrol}

\title{Faster, Further, or Afresh? What AI Can Let Software Developers Unlock}
\title{``To Boldly Go Where No One Has Gone Before'' What Has AI Let Developers Unlock?}

\author{\IEEEauthorblockN{Rudrajit Choudhuri\IEEEauthorrefmark{1},
Jenna L. Butler\IEEEauthorrefmark{2},
Christian Bird\IEEEauthorrefmark{2},
Carmen Badea\IEEEauthorrefmark{2},
Brian Houck\IEEEauthorrefmark{3},
Anita Sarma\IEEEauthorrefmark{1}}
\IEEEauthorblockA{\IEEEauthorrefmark{1}Oregon State University, Corvallis, OR, USA, \{choudhru, anita.sarma\}@oregonstate.edu}
\IEEEauthorblockA{\IEEEauthorrefmark{2}Microsoft, Redmond, WA, USA, \{jennbu, cbird, cabadea\}@microsoft.com}
\IEEEauthorblockA{\IEEEauthorrefmark{3}DX, brian.houck@getdx.com}}


\maketitle

\begin{abstract}
Measuring AI by how much faster developers work is like judging electricity by how much faster it spun a steam-era factory. A fuller measure asks what new work becomes possible.
We surveyed 331 software developers at Microsoft about what AI lets them do that they could not do before, and analyzed their answers using reflexive thematic analysis with a human-in-the-loop, multi-model AI council. Developers described doing familiar work faster (\textit{Faster}, 45\%), undertaking work previously beyond their reach (\textit{Further}, 36.9\%), and taking on new kinds of work centered on building for and directing AI agents (\textit{Afresh}, 12.4\%). 
Across these accounts, learning and scaffolding ran in both directions: developers learned new skills with AI while externalizing their tacit knowledge as the context and standards that guide AI's work. Drawing on Vygotsky's zone of proximal development, we frame developers and AI as joint learners whose zones meet at the skill--judgment nexus: AI can supply skill faster than it can reliably judge its own work, and developers must continuously bridge that gap. Whether they do sets them on one of two paths. On one, developers' judgment grows alongside AI's skill, and they codify what they learn so the loop keeps turning. On the other, they become hollow orchestrators, approving work they can no longer judge and losing the identity and meaning they once drew from it. Based on our findings, we offer a two-way scaffolding playbook to help developers, teams, and tool builders take the first path.
\end{abstract}


\section{Introduction}\label{sec:intro}

Imagine riding a steam train in 1865, the year Jules Verne published a novel about firing three men to the moon from a giant cannon~\cite{verne1865earth}. 
Would you have believed it could be done? Few people on that train would have. A century later, it was an engineering program, and electricity had made it possible.

It did not happen by flipping a switch. When factories first electrified, they swapped the steam engine for an electric motor and ran the same machines in the same layout, and productivity barely moved for decades~\cite{david1990dynamo}. The gains came as people learned what electricity could do and rebuilt their work around it, and as electricity itself was reshaped, from one central dynamo to a small motor on every machine, and later to the electronics that guided a spacecraft. Both the people and the technology had to evolve, and they evolved together.

Software engineering (SE) is at that point with AI. In the 2025 DORA survey, 90\% of tech professionals reported using AI at work~\cite{dora2025}. Yet as a field, we often evaluate AI's impact on developer workflows by measuring how quickly developers complete familiar tasks with AI and how productive they are~\cite{peng2023impact,paradis2025how,cui2026effects}. Like factory owners assessing their first electric motors, we ask how much faster it runs the old work.
Yet, the more interesting question that sits at the horizon is \textit{what AI lets developers do that was out of reach before}, and that horizon is moving quickly. In February 2025, Andrej Karpathy described \enquote{vibe coding}: telling an AI what you want and choosing to \enquote{forget that the code even exists}~\cite{karpathy2025vibe}. Only a year later, he described \textit{agentic engineering}, shifting the emphasis toward orchestrating agents and overseeing their work~\cite{karpathy2026agentic}. Time and output measures alone cannot fully capture this shift, because they do not necessarily reveal what AI makes possible or how it changes developers' work.

Thus, to look at that horizon, we surveyed developers at Microsoft, a company at the frontier of building and using AI systems, investigating:

\begin{itemize}
\item[] \textbf{RQ:} \textit{\textbf{What do software developers report they can now do at work because of AI that they could not do before?}}
\end{itemize}

We analyzed developers' open-ended responses ($N=331$) using a human-in-the-loop, multi-model AI council pipeline following~\cite{choudhuri2026copilot}, guided by reflexive thematic analysis~\cite{braun2006using,braun2019reflecting}. Participants’ accounts fell into three main zones (Section~\ref{sec:rq1}). \textit{Faster} was familiar work, now done faster (45\% of responses). \textit{Further} was work previously beyond their reach or impractical to undertake without AI (36.9\%). \textit{Afresh} was a new kind of work created around AI, such as setting standards for its work, directing agents, and building for them (12.4\%).

To interpret how developers and AI adapt to one another, we draw on Vygotsky's zone of proximal development (ZPD)~\cite{vygotsky1978mind}: the distance between what a learner can do alone and what they can do with guidance from a more knowledgeable other. Yet Vygotsky’s account assumes a reliable guide and a human learner. AI-native SE breaks both assumptions. AI can exceed developers in skills yet fall short in judgment, and it adapts as developers shape its context, memory, and specifications.

We therefore conceptualize developers and AI as joint learners who scaffold each other (Figure~\ref{fig:twolearners}), i.e., developers extend their reach with AI's skill; AI extends its discernment with developers' guidance. We call the meeting point of these two zones the \textit{skill--judgment nexus}. Here, AI can supply skills beyond what it can reliably judge, leaving a gap that developers must continually bridge. In fact, participants described two ways of bridging it: learning with AI, and codifying their own discernment into specifications, memory, and success criteria to guide AI's work. 

We argue that these practices could help developers and their agents grow together. Without them, developers risk becoming \textit{hollow orchestrators}: over-dependent, buried in verification debt, and gradually losing skills and professional identity. Drawing on our findings and the literature, we provide a two-way scaffolding playbook for developers and AI in SE work.

Overall, this paper makes the following contributions: (1) an empirical account of what developers report AI lets them do, across three zones: \textit{Faster}, \textit{Further}, and \textit{Afresh} (Table~\ref{tab:zones}); (2) a conceptualization of developers and AI as joint learners who scaffold each other (Figure~\ref{fig:twolearners}); and (3) the concept of the \textit{skill--judgment nexus} and a two-way scaffolding playbook that adapts literature based scaffolding functions that can support developer--AI learning and keep the loop turning (Table~\ref{tab:playbook}).

\section{Related Work}\label{sec:related}

As AI becomes native to software development, an extensive line of research has examined how AI affects developers' productivity and experiences~\cite{peng2023impact,cui2026effects,butler2025dear}, how developers trust and adopt these tools~\cite{choudhuri2025trust,choudhuri2025needs,russo2024navigating}, and where they want its support in their daily work~\cite{khemka2024toward,choudhuri2026ai,choudhuri2026copilot}.

Productivity findings vary across tasks and settings.
For example, controlled trials report developers completing tasks 55\% faster with AI~\cite{peng2023impact} and spending about 21\% less time on it~\cite{paradis2025how}, alongside field experiments finding that developers could now complete 26\% more tasks~\cite{cui2026effects}. In contrast, Becker et al.~\cite{becker2025measuring} found that developers working in their own repositories took 19\% longer with AI, while believing it had sped them up.

Studies of everyday AI use show where developers seek support and what effort that support requires. Khemka and Houck~\cite{khemka2024toward} found that developers most wanted AI to take on repetitive work such as generating tests and writing documentation, while fewer wanted it to write new code, the work that gave them a sense of accomplishment. Yet reported gains in speed and output can coexist with increased review burden and persistent verification effort~\cite{ choudhuri2026copilot, afroz2026fast}. Mozannar et al.~\cite{mozannar2024reading}, for instance, found that verifying suggestions took the largest share of session time, and Choudhuri et al.~\cite{choudhuri2025needs} found that when output drifted from developers' goals, they spent more effort prompting, verifying, and modifying it, which eroded their trust and adoption of these tools. He et al.~\cite{he2026speed} found that short-term velocity gains following Cursor adoption were accompanied by increased code complexity and static analysis warnings, which contributed to later slowdown.

A complementary line of work investigates not only whether AI \textit{can} support developers' work but whether it \textit{should}, and to what extent. For instance, Choudhuri et al.~\cite{choudhuri2026ai} examined how developers' cognitive appraisals of their daily work predicted their openness to and use of AI across software development tasks: developers sought AI to refine their craft in core work and to reduce toil, and limited it in identity-defining and relationship-centric work. Subsequent work examined how the meaning developers assign to their work helps predict where they draw the line on AI autonomy~\cite{choudhuri2026you}, and which systems they want built for this AI-native era of knowledge work~\cite{choudhuri2026copilot}. However, these studies primarily take existing work as their frame of reference, asking which of today's tasks AI can and should support. In this study, we focus instead on what AI has let developers do that they could not do before.

Notably, as AI becomes more autonomous, developers' work itself is changing, from writing code toward directing and checking it~\cite{choudhuri2026ainative}. The shift reaches beyond individual workflows: AI-native startups already run with teams about 25\% smaller and flatter than comparable startups~\cite{kim2026ainative}. Within a developer's own work, vibe coding shifts expertise toward managing context and evaluating output~\cite{sarkar2025vibe}, although participants working with an autonomous agent reported understanding its work less well than non-agentic AI~\cite{chen2026code}. Taking on more work with AI thus also changes what developers must understand and judge. These responsibilities reflect the metacognitive demands AI places on its users, such as specifying goals, evaluating output, and deciding when and how to use it~\cite{tankelevitch2024metacognitive}.

Taken together, these perspectives show how AI affects current development workflows. Our study extends this work by examining how AI has enabled developers to do what they could not before, providing a snapshot of their reported experiences as of mid-2026. We distinguish \textit{Faster}, \textit{Further}, and \textit{Afresh}: familiar work accelerated, previously unreachable work made possible, and new work created around AI.

To interpret how developers and AI co-evolve in a shifting landscape, we draw on Vygotsky's ZPD~\cite{vygotsky1978mind} and conceptualize developers and AI as joint learners who scaffold each other~\cite{wood1976role}. Finally, we present a two-way scaffolding playbook to support AI-native SE work as developers' skills and AI's capabilities continue to evolve.
\vspace{-2mm}
\section{Method}\label{sec:method}

Our goal was to understand what AI enables developers to do that they could not do before. To investigate this, we surveyed software developers at Microsoft, spanning diverse domains, teams, stakeholder contexts, and geographies. Developers there routinely use both mature and emerging AI tools, making it a rich and diverse setting for our study. Microsoft’s institutional review board approved our study.

\subsection{Survey}\label{sec:survey}

Our data come from a survey of Microsoft developers fielded in August 2026 as a follow-up run to the \textit{AI Where It Matters} study~\cite{choudhuri2026ai}.
One goal of the survey was to understand how AI is expanding the reach of work developers perceive themselves as able to do, and what these new capabilities may reveal about the future of software work. To examine this, the survey included an optional open-ended question: 

\textit{Is there anything you can now do at work because of AI that you couldn't do before? If so, what is it, and what has it enabled you to do?} 

We intentionally left the question broad, so that developers could identify for themselves what counted as a newly accessible capability. Responses to this question form the paper's primary unit of analysis.

\textbf{\textit{Data collection and participants.}} We administered the survey in Qualtrics and invited 4,000 software developers at the company, sampled uniformly at random across product groups, experience, domains, and geographies. We sent one reminder after one week and received 355 responses (8.9\% response rate, consistent with prior SE surveys~\cite{kalliamvakou2019manager, choudhuri2026ai}). Of these, 345 contained substantive answers. We retained 331 participant responses for analysis after excluding 14 whose code assignments lacked sufficient support, and refer to these participants as P1--P331. 

Table~\ref{tab:participants} summarizes the sample. Participants spanned a wide range of industry experience, from $\le$ 5 years (23.3\%) to more than 20 years (23.6\%), and 77\% self-identified as men. AI was already well integrated into their work: 89.1\% reported that it helped them \emph{a lot} or \emph{quite a bit} in their daily work.

\begin{table}[tp]
\caption{Participant demographics (N = 331)}
\vspace{-2mm}
\label{tab:participants}
\scriptsize\centering\setlength{\tabcolsep}{5pt}
\begin{tabular}{@{}ll>{\raggedleft\arraybackslash}p{0.8cm}>{\raggedleft\arraybackslash}p{0.8cm}@{}}
\toprule
 & & $n$ & \% \\
\midrule
Years in industry & 0--5 & 77 & 23.3 \\
 & 6--10 & 72 & 21.8 \\
 & 11--15 & 56 & 16.9 \\
 & 16--20 & 48 & 14.5 \\
 & Over 20 & 78 & 23.6 \\
\addlinespace[2pt]
Gender & Man & 255 & 77.0 \\
 & Woman & 69 & 20.8 \\
 & Non-binary / diverse & 1 & 0.3 \\
 & Prefer not to say or blank & 6 & 1.8 \\
\addlinespace[2pt]
AI helps get & A lot & 183 & 55.3 \\
work done & Quite a bit & 112 & 33.8 \\
 & Somewhat & 29 & 8.8 \\
 & A little & 7 & 2.1 \\
 & Not at all & 0 & 0.0 \\
\bottomrule
\end{tabular}
\vspace{-5mm}
\end{table}

\subsection{Data analysis}\label{sec:analysis}

We analyzed developers' open-ended responses using a human-in-the-loop, multi-model council pipeline following~\cite{choudhuri2026copilot}, guided by reflexive thematic analysis~\cite{braun2006using,braun2019reflecting}. We leveraged the pipeline to make large-scale qualitative analysis tractable without delegating interpretive authority to the models~\cite{tai2024examination}. Specifically, models performed bounded, auditable operations: surfacing candidate patterns together with their evidence and, once the codebook was fixed, applying those codes to the corpus. Two authors retained responsibility for what the patterns meant, where their boundaries lay, and which interpretations entered the final analysis.

Our council comprised three frontier models from distinct provider families: Claude Opus 5.5 (Anthropic), GPT-5.6 (OpenAI), and Gemini Flash 3.7 (Google). Using different model families reduces the risk that a single model's inductive biases or failure modes dominate the analysis~\cite{wu2026council}; agreement across families also provides stronger convergent evidence than agreement within one family, where shared training data and optimization can produce shared errors~\cite{bommasani2021opportunities,liang2022holistic}. Throughout the process, every proposed pattern and coding decision remained traceable to participants' verbatim responses. Specifically, we proceeded as follows:

\textbf{\textit{Stage 1: Independent theme discovery.}} The three models independently analyzed the complete set of answers, with no visibility into one another's output. We gave each model three analytic lenses: \emph{semantic}, focused on what participants stated explicitly; \emph{latent}, focused on assumptions underlying those statements; and \emph{contrastive}, focused on differences across responses. Each model proposed candidate themes, sub-themes, and their codes, and supplied the participant IDs and verbatim excerpts supporting them. We required each candidate to be grounded in at least three distinct participants. 


\textbf{\textit{Stage 2: Codebook reconciliation.}} We then reconciled the independently discovered candidate sets using GPT-5.6. The model was instructed to (1) merge candidates that described the same underlying concept, (2) retain candidates proposed by only one model only when supported by at least three distinct participants, and (3) remove candidates that were not grounded in concrete developer experiences or were too sparsely supported. We used a single model at this stage because the task was consolidation rather than discovery: the themes and codes had already been surfaced independently by three model families, and reconciliation required organizing that evidence into a coherent set for author review.
Each reconciled theme and code kept a record of which models proposed it and under what original labels, for author review.

\textbf{\textit{Stage 3: Author review and codebook approval.}} Two authors then reviewed every candidate theme and code against all responses cited as evidence. For each, they assessed three properties: \emph{fidelity}, whether the code captured what developers described; \emph{grounding}, whether the cited responses genuinely supported that interpretation; and \emph{distinctness}, whether it was sufficiently distinct for systematic coding. The authors documented positive and negative examples to make those boundaries explicit, then compared their assessments through negotiated agreement. Through iterative discussion, they split overly broad codes, merged redundant ones, added missing distinctions, and removed interpretations that were insufficiently grounded. 
Systematic coding began only after the authors approved the codebook.

\textbf{\textit{Stage 4: Systematic coding and independent verification.}} With the codebook fixed, three new model instances (Claude Opus 5.5, Gemini Flash 3.7, GPT 5.6) independently coded every participant response. Each model was instructed to provide a rationale before assigning codes~\cite{wei2022chain}, leaving an auditable link between the response and the assigned code. Models could assign multiple codes when an answer expressed multiple ideas, but they could not introduce new codes during this stage, preventing code drift.

\textbf{\textit{Stage 5: Agreement and consensus.}} We assessed consistency among the three model coders using Krippendorff's $\alpha$ for each code's binary assignment~\cite{krippendorff2018content}. Across codes, the median $\alpha$ was 0.92 (min = 0.83, max = 0.97). We use these values as a diagnostic of coding consistency, rather than as a substitute for author interpretation. Final code assignments required a 2-of-3 majority, applied independently to each response and code. 

We then subjected the resulting assignments to an independent verification layer. We used Typesafe's Jev~\cite{typesafe2026jev}, a calibrated decision model that returns judgments with probabilities, to independently evaluate each response--code link and each response's zone assignment (described below). Across all, Jev assigned a median probability of 0.84 (IQR: 0.78–0.93). It also assessed whether every pair of sub-themes within a theme was meaningfully distinguishable. 


As a further robustness check, one author independently re-coded a random sample of 50 responses (15\% of the corpus) using the final codebook, without seeing the council's assignments. Human--council agreement was $\kappa = 0.87$, providing an independent check on the systematic coding.

\textbf{\textit{Zone assignments.}} Finally, we mapped the themes and codes to three zones based on what developers reported AI had enabled them to do (see Table~\ref{tab:zones}). \textit{Faster} captured work developers could already do, but could now do faster or at greater volume with AI. \textit{Further} captured work that became reachable with AI, either because developers had not previously undertaken it or because it required additional knowledge or skills; broader scope, coverage, and coordination across projects or teams also fell in this zone. 
\textit{Afresh} captured work that emerged around developing and directing AI-enabled systems, such as defining the context, standards, and expectations that should govern AI's work, directing agents, or building systems and pipelines around them.
We excluded 14 responses whose zone placement lacked sufficient support (recall Sec.~\ref{sec:survey}).

\textbf{\textit{Reflexivity note.}} We used AI to extend the scale of qualitative analysis rather than to replace its interpretive work: models proposed candidate patterns, applied an author-approved codebook, and flagged ambiguous cases, while the authors decided what each pattern meant and how patterns should be synthesized. Every finding remains grounded in participants' responses, which readers can inspect in Section~\ref{sec:rq1}.

\textbf{Supplemental.} The analysis prompts, codebook, and pipeline scripts are available in the supplemental material~\cite{choudhuri2026supplemental}.

\subsection{Threats to validity}\label{sec:threats}

\textit{\textbf{Construct validity.}} We report stated experience rather than observed performance. Developers may describe gains they would not realize in practice, and self-reports of speed can be wrong even in direction~\cite{becker2025measuring}. We treat this as motivation for future work: the three zones provide testable categories for measuring what AI changes in developers' work.

\textit{\textbf{AI-assisted analysis.}} Using AI as coders risks surfacing plausible themes that do not fully reflect what participants meant. We addressed this at each stage: three models discovered themes independently under distinct lenses, anchoring each in participant IDs and verbatim quotes; two researchers reviewed every candidate against its cited responses before approving the codebook; models gave a rationale for every code assignment; and a decision model and blind author re-coding of a random sample verified the assignments. The verbatim quotes throughout the paper also let readers judge whether the themes are grounded in participants' responses.

\textit{\textbf{Internal validity.}} Ours is a cross-sectional study that captures developers' stated experience at a single point in time. As AI tools and developer experiences co-evolve, the zones provide criteria for tracking shifts in what developers can do with AI, including whether \textit{Further} work becomes \textit{Faster} or \textit{Afresh} work becomes routine. Self-selection remains a potential confound: respondents may hold stronger views about AI support than the broader population.

\textit{\textbf{External validity.}} 
We studied Microsoft developers across global sites, diverse teams, domains, and processes. 
Microsoft may provide developers with access to capabilities and organizational support that are less readily available in other settings, potentially influencing the types of AI-enabled work they report. The capabilities we identify reflect developers' accounts of daily work, though their salience may differ among less frequent AI users, in smaller organizations, in open-source settings, or in regulated industries. We therefore present an in-depth account of a large organizational context rather than a claim to represent all software developers. Single case studies, like ours, have advanced scientific discovery~\cite{flyvbjerg2004five} and produced insights in software engineering~\cite{kalliamvakou2019manager,miller2026examples}. Replication in other contexts remains necessary future work.

\section{Results}\label{sec:rq1}

Participants described AI as changing both the pace and scope of their work. Their accounts included familiar tasks completed faster, previously deferred or inaccessible work brought within reach, and new responsibilities arising from the use of agents. One participant illustrated the scope of what they could undertake: \pq{AI lets me move from idea to a working, tested prototype far faster, so I spend my time on architecture and judgment instead of code, and I can take on org-scale problems that would have been out of reach solo} (P174).

We grouped participants' responses into three zones. \textbf{\textit{Faster}} comprised work participants could already do alone but completed faster with AI (45\% of responses). \textbf{\textit{Further}} comprised work AI brought within reach by providing skills, knowledge, or capacity participants lacked, or by making previously impractical efforts worthwhile (36.9\%). \textbf{\textit{Afresh}} captured new work arising from AI: directing agents, setting standards for their work, and building for them (12.4\%). The remaining 5.7\% of responses named no specific task and were not assigned to a zone. Table~\ref{tab:zones} summarizes the findings with representative scenarios from the data. We describe each zone below.

\begin{table*}[tp]
\caption{What AI has let developers unlock: zones, themes and sub-themes, with representative scenarios from the data. A theme's percentage is the share of total responses (331) that carry any of its sub-themes. Responses can carry several themes, so theme percentages overlap and do not sum to the zone total, which counts each response once.}
\vspace{-2mm}
\label{tab:zones}
\fontsize{7.2}{8.2}\selectfont\centering\setlength{\tabcolsep}{3pt}\renewcommand{\arraystretch}{0.93}
\begin{tabular}{@{}>{\raggedright\arraybackslash}p{0.145\textwidth}>{\raggedright\arraybackslash}p{0.225\textwidth}>{\raggedright\arraybackslash}p{0.61\textwidth}@{}}
\toprule
\textbf{Theme} & \textbf{Sub-theme} & \textbf{Representative scenarios from the data} \\
\midrule
\rowcolor{zfaster}[\tabcolsep][0pt]\multicolumn{3}{@{\hspace{3pt}}l@{}}{\textbf{Faster: the same tasks, faster} (45.0\%)} \\
\addlinespace[1pt]
Familiar work, faster (35.6\%) 
 & Code and delivery acceleration & Writing code much faster, in any language; Several comprehensive pull requests a day; Scripts coded up quickly; Queries and dashboards generated; Older functions refactored in days \\
\arrayrulecolor{black!18}\cline{2-3}\arrayrulecolor{black}
& Exploration and planning acceleration & Brainstorming design ideas and trade-offs faster; Planning and prioritizing upfront\\
\arrayrulecolor{black!18}\cline{2-3}\arrayrulecolor{black}
 & Debugging and diagnosis acceleration & Root-cause analysis under pressure; Troubleshooting production issues much faster; Resolving bugs without asking others; Security fixes implemented much faster \\
\arrayrulecolor{black!18}\cline{2-3}\arrayrulecolor{black}
 & Testing and review acceleration & Test cases and thorough testing done in minutes; Faster code and pull-request review \\
\arrayrulecolor{black!18}\cline{2-3}\arrayrulecolor{black}
 & Docs and reporting acceleration & Technical documentation drafted in minutes; Faster presentations and decks; Stakeholder and cross-functional updates; Design documents and specifications; Weekly and status reports \\
\arrayrulecolor{black!18}\cline{2-3}\arrayrulecolor{black}
 & Generic acceleration & A much faster pace of work; Less time per task; Faster deliveries and project completions \\
\cmidrule(l){1-3}
Automating toil and delegating work (21.1\%) & Toil automation & Repetitive manual tasks and daily workflows; Rebasing and smoke testing; Flaky-test fixes and cleanup at volume; Boilerplates; Administrative chores; Work items and pull-request descriptions \\
\arrayrulecolor{black!18}\cline{2-3}\arrayrulecolor{black}
 & Background work running in parallel & Lower-priority coding tasks left to an agent in the background; Coding continuing while in meetings; Multiple agents at once from multiple dev boxes; Investigations running while other work continues \\
\arrayrulecolor{black!18}\cline{2-3}\arrayrulecolor{black}
 & Delegating structured work to AI & Well-described work items, analysis, and debugging handed to AI agents; Agents analyzing logs and generating root-cause hypotheses; Implementation plans prepared by agents \\
\cmidrule(l){1-3}
Comprehension and learning, faster (35.3\%) & Unfamiliar codebase comprehension & An unfamiliar codebase understood in hours; Searching large codebases; Grokking legacy code; Ramping up on new projects and repositories; Onboarding to a new team faster\\
\arrayrulecolor{black!18}\cline{2-3}\arrayrulecolor{black}
 & Information synthesis & Researching technical questions faster; Finding the relevant documentation or section; Summarizing long documents and technical content; Summarizing email, chats, meetings and recordings \\
\arrayrulecolor{black!18}\cline{2-3}\arrayrulecolor{black}
 & Learning new domains faster & Tailored self-directed learning; AI setting up learning steps and challenges; Learning new domains faster; Asking instead of reading long documentation \\
\arrayrulecolor{black!18}\cline{2-3}\arrayrulecolor{black}
 & Large-scale data interpretation & Ad-hoc analysis of large datasets without a script; Pulling data together from several sources; Analyzing large volumes of logs and telemetry \\
\midrule
\rowcolor{zfurther}[\tabcolsep][0pt]\multicolumn{3}{@{\hspace{3pt}}l@{}}{\textbf{Further: tasks reachable only with AI} (36.9\%)} \\
\rowcolor{zfurtherB}[\tabcolsep][0pt]\multicolumn{3}{@{\hspace{3pt}}l@{}}{\textit{Tasks that needed someone else's knowledge or a skill they lacked} (24.2\%)} \\
\addlinespace[1pt]
Working beyond one's own skill, knowledge, or capacity (24.2\%) & Development work outside one's expertise or role & Work in unknown frameworks, languages, and domains; Deployment pipelines without prior skills; Front-end work by a back-end engineer \\
\arrayrulecolor{black!18}\cline{2-3}\arrayrulecolor{black}
 & Answering questions that needed SMEs & Understanding a feature without asking SMEs; Troubleshooting without bothering one's boss or peers \\
\arrayrulecolor{black!18}\cline{2-3}\arrayrulecolor{black}
 & Work beyond one person's capacity & Having a team of agents at disposal; Work that would have taken two or more engineers; Patterns across more meeting transcripts than one person can hold \\
\arrayrulecolor{black!18}\cline{2-3}\arrayrulecolor{black}
 & Communicating beyond one's own skill & Writing clearer English than one could alone; Better presentation of one's own thoughts; Better feedback for direct reports; Analyzing one's own communication \\
\arrayrulecolor{black!18}\cline{2-3}\arrayrulecolor{black}
 & Coordinating contexts and projects across teams & Managing several parallel teams; Managing more projects and products than before\\
\midrule
\rowcolor{zfurtherA}[\tabcolsep][0pt]\multicolumn{3}{@{\hspace{3pt}}l@{}}{\textit{Tasks they could do but never started} (12.7\%)} \\
\addlinespace[1pt]
Deprioritized work now feasible (9.7\%) & Finally getting to backlog work & Features put off as months of work; Backlog items and small enhancements; Technical debt as a side project\\
\arrayrulecolor{black!18}\cline{2-3}\arrayrulecolor{black}
 & Building personal tools and automations & Scripts for one's own workflows; Building software and visualization tools for one's own workflows; A tool tying team objectives to tracked features; Vibe-coded personal and quality-of-life tools \\
\cmidrule(l){1-3}
Expanded scope and rigor (6.3\%) & Work scope expansion & Bigger refactors and larger projects; Broader, more complex problems, ideas, and tasks taken on\\
\arrayrulecolor{black!18}\cline{2-3}\arrayrulecolor{black}
 & Testing and rigor expansion & Large volumes of unit tests; Proofs and tests that check one's own hypotheses; Deep analysis and test coverage of all code branches \\
\cmidrule(l){1-3}
Expanded exploration (7.9\%) & Exploring with prototypes & Quick proofs of concept(s); Prototyping without committing to a solution; Working prototypes to get an idea across to another team; Greenfield prototypes not meant to ship \\
\arrayrulecolor{black!18}\cline{2-3}\arrayrulecolor{black}
 & Comparing approaches before committing & Checking multiple approaches before committing; Analyzing several design approaches with trade-offs\\
\midrule
\rowcolor{znew}[\tabcolsep][0pt]\multicolumn{3}{@{\hspace{3pt}}l@{}}{\textbf{Afresh: new work because of AI} (12.4\%)} \\
\addlinespace[1pt]
Directing and supervising agents (6.3\%) & Directing and managing agents & Orchestrating work streams rather than executing steps; Teams of agents managed like teams of engineers; Most code written by AI; Conversing with AI, even by voice, instead of writing the artifact; Barely opening an editor except to diff \\
\arrayrulecolor{black!18}\cline{2-3}\arrayrulecolor{black}
 & Maintaining oversight of agents & Oversight of AI-generated work before it merges; Oversight on agents carrying work from issue to PR; Agents that escalate only consequential decisions; Agents that notify and halt when exit criteria can't be met \\
\cmidrule(l){1-3}
Setting the standard for AI's work (2.7\%) & Specifying intent, tests and exit criteria & Evaluations against success criteria; Automated validation of AI-generated changes; Defining intent and reviewing the agent's reasoning \\
\arrayrulecolor{black!18}\cline{2-3}\arrayrulecolor{black}
 & Setting up a council of models (collaborative and adversarial) & Designs passed through different agents and models; Agent orchestrators that plan, code, and review each other's work; Models arguing different points of view \\
\cmidrule(l){1-3}
Building for agents (4.8\%) & Building agents, harnesses and pipelines & Agent-driven pipelines and custom skills; Personalized agents; A custom harness with evaluations; Agents that investigate telemetry\\
\arrayrulecolor{black!18}\cline{2-3}\arrayrulecolor{black}
 & Writing context and memory for agents & Specifications for agents; Context engineering for long workflows; Durable memory and a shared backlog \\
\bottomrule
\end{tabular}
\vspace{-4mm}
\end{table*}

\subsection{Faster: the same tasks, faster}\label{sec:faster}

When asked what AI let them do that they could not do before, many participants described a change in pace rather than the fundamental nature of their work: \pq{I could do everything I currently do without AI; however, I can do my work faster.} (P5). Many reported gains in terms of saved time. For instance, one described \pq{drafting a 90-page document in 20 minutes and completing it, including review, in two days instead of two weeks} (P59). Another reported completing \pq{a yearly project in 2 months from blueprint to production release} (P32).

\textbf{Familiar work, faster.} Participants reported gains across familiar activities. For example, during design, AI helped them explore alternatives: \pq{AI quickly generates multiple ideas and explains the pros and cons of each} (P186). In implementation, one participant could \pq{complete two relatively comprehensive PRs in a day. Previously, even simple features took 1-2 days to develop, build, and test.} (P184). Similarly, testing sped up, with \pq{strong testing done in minutes} (P159), and debugging \pq{once a huge task, was now a couple of prompts away} (P250).

\textbf{Automating toil and delegating work.} Participants handed AI repetitive tasks and structured portions of engineering work. They kept the portions that needed their attention and ran the rest in parallel: \pq{I get a lot more low-priority tasks done (in the background) while I work on high priority portions in the foreground} (P6). Some reorganized their workflow around AI's latency: \pq{I used to only do one thing at a time. But now I will do 2 or 3 while waiting for AI to respond on each problem} (P274). Agents extended this pattern, allowing work to continue while developers attended to other responsibilities: \pq{I give a task to the agent, which will work on it while I handle other tasks (like meetings) and I'll have results to review} (P320). The overall payoff showed up at the scale of the workday: \pq{They may not get done faster in isolation, but more work lands in that timeframe} (P259).

\textbf{Comprehension and learning, faster.} While prior work warns that AI can let developers produce code they do not understand~\cite{shen2026how,prather2024widening}, participants in our study described using AI at an earlier point in their work to find relevant information and build project understanding: \pq{Instead of spending days going through documents, code, and multiple KT sessions, I can use AI to quickly understand the context and identify what matters most} (P247). This, in fact, lowered the barrier to entering unfamiliar codebases: \pq{I can work on new codebases within 30 minutes, when it would take me multiple days beforehand to get acquainted} (P11). They described using the time saved for work requiring engineering judgment: \pq{This lets me spend more time on design decisions, validating solutions, and solving higher-value engineering problems} (P178).

Participants also noted substantive learning gains: \pq{Learning outside of my domain is significantly faster due to tailored responses. I can quickly understand a topic and go in depth} (P80). Others used AI to structure the learning process itself: \pq{using AI to set up a set of learning steps and challenges. It's a way of learning that wasn't feasible pre-AI.} (P203).



\subsection{Further: tasks reachable only with AI}\label{sec:further}

One participant described contributing to front-end development, \pq{something I previously had limited opportunities to do because I did not have enough time to learn React and other front-end technologies} (P140). Another could now \pq{implement features in hours that were previously put off due to being weeks to months of work} (P107). In 36.9\% of responses, participants described such instances. Most of it previously required skills, knowledge, or capacity participants lacked (24.2\%), and was work they could have done but had previously deferred because of the effort required (12.7\%).

\textbf{Working beyond one's own skill, knowledge, or capacity.} Participants noted substantive horizontal skill expansion. These ranged from deployment pipelines \pq{which I did not have the skills to do before} (P30) to data analysis and reporting for which they had been \pq{fully dependent on others with skills in SQL, data science, etc.} (P90). Broader skills also let participants take responsibility for larger portions of a system: \pq{I can cover E2E now. Previously, I was limited to only the backend} (P28). These gains also extended to interpersonal work: \pq{I can write better feedback for people and [it] helps me think through how to work with my directs} (P12).

Accessible knowledge through AI also reduced the social cost of seeking help~\cite{lee2002social}: \pq{I can learn and troubleshoot without bothering my boss or my peers} (P151). Capacity expanded as well. Some could now pursue more of their own ideas than they could staff alone: \pq{AI gives me a virtual team to go explore all the ideas I have and magnify my impact} (P199).

\textbf{Deprioritized work now feasible.} Some work had always been discretionary: worth doing, but only when time allowed. AI made it cheap enough to do.
For instance, one participant had AI \pq{write a script to do tedious updates on web portals that don't expose APIs and otherwise require manual browser navigation. I recently automated it in an hour that saved me 80 hours of manual work. I never would have done those 80 hours, so those items would never have been updated. Writing the automation without AI would have probably taken me 30 hours, which would still have been too hard to justify the investment.} (P197) Technical debt followed suit: \pq{I'm able to rewrite older functions in days and modernize our code base as a side project. Earlier, we used to have to dedicate specific time and resources to it} (P223).

Personal software became worth building, too: \pq{Tools, visualizations, workflow automation highly personalized to my needs} (P327). Small, personal-scale problems finally became worth solving: \pq{I can now solve more side problems/automate personal-scale solutions and workflows. There are many quality-of-life issues before that would have taken too much time to improve, that now we can solve on the side with a vibe-coded tool} (P269). Some began planning projects they would never have considered before: \pq{I have side projects I'm looking forward to spending time on that I would've never considered without AI} (P17).

\textbf{Expanded scope and rigor.} Participants described taking on more work and being able to do it more thoroughly: \pq{It feels like I'm able to take up broader and more complex tasks, go in greater breadth AND depth \ldots\ [and] deliver higher quality work} (P74). 
Some participants recounted evaluating their own work more rigorously, building evidence for and against their own ideas: \pq{I can generate proofs, including tests, that validate and prove hypotheses \ldots\ Before, this was very difficult to do as I was already biased to my own solution} (P134).

\textbf{Expanded exploration.} A lower cost per attempt changed how participants explored options and made decisions. Proofs of concept(s) became cheap and disposable, even in unfamiliar technologies. Participants noted that they could now evaluate ideas through working implementations: \pq{I can just implement my ideas easily and see how they work instead of speculating and changing the idea if it doesn't work} (P79). Prototypes also brought stakeholder feedback forward: \pq{I can build rapid, useful prototypes within a few hours \ldots\ and obtain quick feedback from the stakeholders; whereas, previously, it would have taken a few weeks to work on a wireframe concept} (P267). Furthermore, cheaper exploration changed when participants committed to a solution: \pq{the cost of investigating a hypothesis dropped enough that I check three or four approaches before committing, where I'd previously have committed to the first plausible one} (P158).



\subsection{Afresh: new work because of AI}\label{sec:new}

As AI took on more of the engineering work, a new kind of work grew around it: directing AI, deciding when its work was done, and building what it needs (12.4\% of all responses).

\textbf{Directing and supervising agents.} For some participants, AI changed the nature of the job role itself: \pq{my role has shifted from executing every step myself to orchestrating work streams and evaluating outcomes.} (P116). In fact, orchestration took several forms. Some noted workflows that allowed \pq{teams of agents to be managed like teams of engineers} (P300). For some, the editor had become primarily a place to inspect changes: \pq{I barely open an editor except for diffing and minor corrections.} (P294). Others described producing artifacts through iterative conversations with agents: \pq{Instead of actually writing an artifact (code, test, design doc, ppt), I talk with my agents and keep iterating until I am satisfied} (P309); \pq{I use a voice-based prompt and have the response read aloud so you can interact with the agent and get things done} (P288). In one team, AI already wrote most of the code: \pq{Now, more than 90\% of the code is AI-written. We act as a protective layer, as a manager of agents.} (P114).

Managing agents also meant taking responsibility for their output. Participants stressed reviewing AI-generated work before accepting it: \pq{In the end, I'm the gatekeeper of my development work and I take responsibility for AI-generated [work] prior to it merging into our branches.} (P303). As agents handled broader workflows, oversight included checking whether they preserved specifications and intent across stages: \pq{I use AI to drive my entire workflow, from ticket creation based on PRD, design documents, development and testing automation, PR reviews, telemetry and live site. I review whether it has carried the specs and intent from one phase to another} (P311). Others envisioned more selective intervention, expecting agents to \pq{only escalate when consequential decisions or evidence-backed blockers needed resolution} (P298).

\textbf{Setting the standard for AI's work.} Directing agents raised a new question: \textit{when is the work considered done?} AI by itself did not necessarily provide a reliable criterion: \pq{Right now AI sort of just stops. There's some probabilistic mechanism that determines it is time to await a human reply} (P295). Participants thus described building harnesses that checked output against artifacts specifying the finished result. 

Some expected defining these standards to become a larger part of their daily work: \pq{I foresee that most of engineering will become about iterating on the success and exit criteria more than the artifacts implemented to meet them} (P39). For others, the shift had already happened: \pq{my role shifts from doing the steps to defining intent and reviewing the agent's reasoning and results} (P296). These participants noted that testing and reviewing agents' work consumed much of their time, leading them to seek ways to automate parts of it: \pq{my time is mostly for testing and reviewing the work that is done by agents. So, I am seeking ways to automate testing agentic workflows} (P317). Additionally, a few enlisted agents as critics, \pq{passing designs through AI over and over for review (different agents, different models)} (P301), or used \pq{AI orchestrators that plan, code, \& review each other's work} (P330).

\textbf{Building for agents.} Participants built the context and infrastructure that agents needed to work. Part of this was shaping the material agents read and the memory available to them. Longer workflows also called for efficient context management: \pq{It's all about the context; we need a way to manage that better, especially for longer thread workflows - this means changing the way we write docs.} (P321). Specifications were already changing: \pq{Dev and PM specs are increasingly using the Markdown file format so AI can read them.} (P331).

Participants also explored ways to structure execution and maintain memory. One suggested that explicit pipeline definitions could bring \pq{certainty where now we have probability}, and had explored a \pq{\enquote{dreaming} phase over memories that curates existing and generates new insights} (P326). Another wanted to build a pipeline for agents to \pq{coordinate through a shared backlog and durable memory} (P298).

For some, the infrastructure around the model was becoming a product in its own right: \pq{harnesses \textit{as platforms}} (P329), alongside software products designed for agents to use. The volume of AI-generated changes also created a need for tools to make them easier to review: \pq{I'm working on an AI tool to break down large code diffs (which are increasingly common now that people can generate a 90-file PR) into smaller logically-grouped chunks} (P316).

Beneath the harnesses, pipelines, and review practices lay a familiar challenge: \textit{knowing when trust was warranted}. For some participants, trust itself became the new design requirement: \pq{changing how the AI report[s] back to a sufficient trust level, and changing how managers [of agents] can have enough affordances to verify their trust} (P300).


\section{Discussion}\label{sec:discussion}

\begin{figure*}[tp]
\centering
\includegraphics[width=0.88\textwidth]{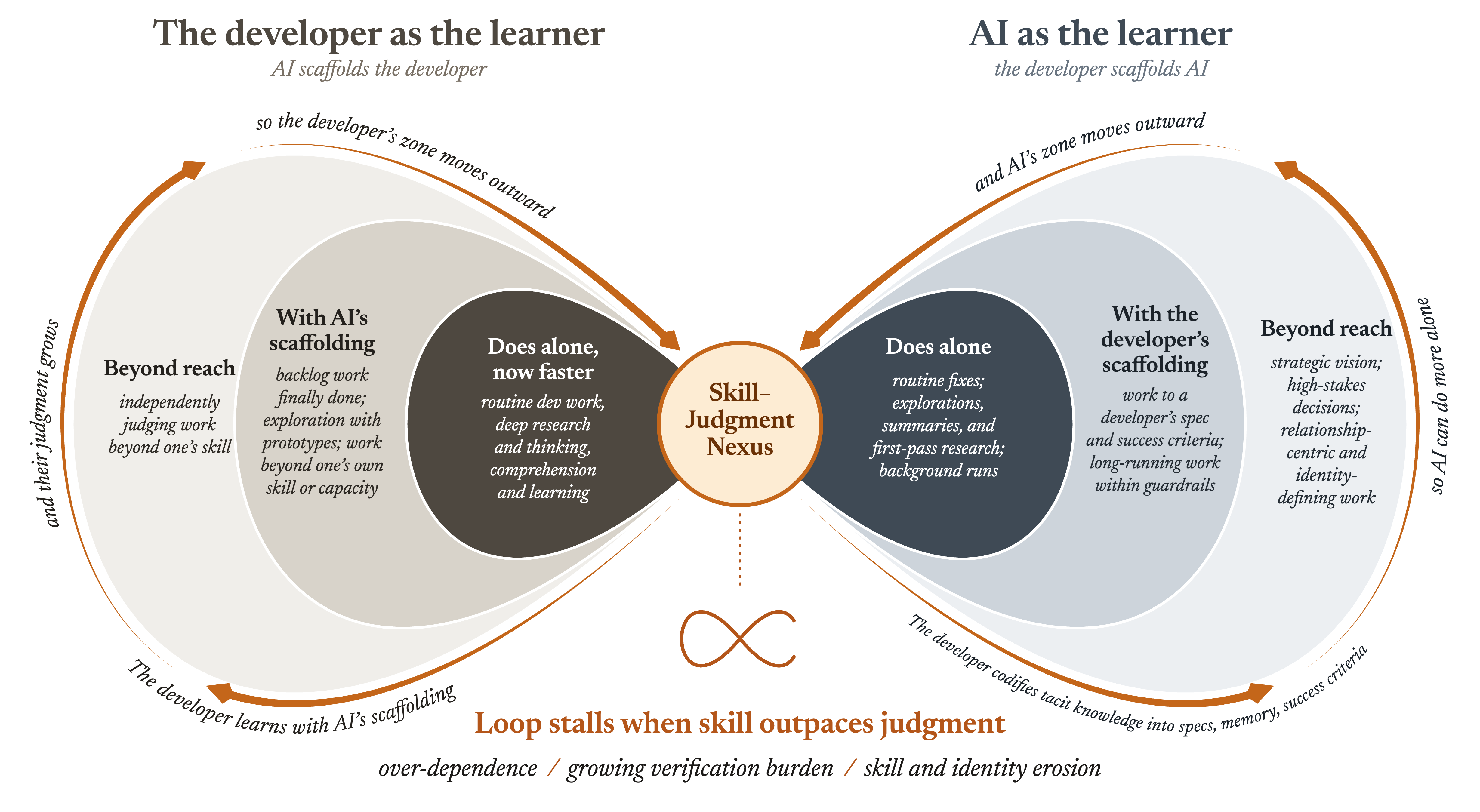}
\caption{\small Developers and AI as joint learners in the zone of proximal development. Each side shows what the learner does alone, with scaffolding, and what is beyond its reach. The zones meet at the skill--judgment nexus: the loop turns when developers grow their judgment and codify it for AI, and stalls when skill outpaces judgment.}
\label{fig:twolearners}
\vspace{-4mm}
\end{figure*}

Our findings suggest that developers now play two roles in an AI-native workplace. As learners, they use AI to go further and to start afresh, taking on work they could not have done alone or would never have attempted. As teachers, they pass on to AI what years of practice have taught them, through the specifications, success criteria, and memory that shape what AI agents can do. A developer can therefore be a learner and a teacher within the same workday.

Yet neither role stands still. What AI can do alone, what it still needs a developer for, and what a developer must know to judge its work all shift each time a model improves, each time a developer externalizes what they know, and each time a new kind of task appears. Today's skills, therefore, matter less than the motion. Consider an agent that a team has worked with for a year. It is no longer the system they installed, because the people around it have rewritten its instructions and tests many times. They now spend their days differently because of what it took on. Engelbart described augmentation this way six decades ago: people, language, methods, and artifacts form one system, and a change in any part, including the tools, calls for a change in the people~\cite{engelbart1962augmenting}. Lehman made the same observation about software: a program used in the real world must keep changing or become less useful~\cite{lehman1980programs}.

\subsection{Developers and AI as joint learners, meeting at the skill--judgment nexus}\label{sec:twolearners}

To make sense of this continuing change, we draw on Vygotsky's zone of proximal development~\cite{vygotsky1978mind}, a theory of guided learning. In this view, every learner works at three distances: what they can do alone, what they can do with help from a more knowledgeable other, and what lies beyond them even with help. Scaffolding is the help that makes the middle distance reachable. Good scaffolding starts from what the learner can already do, adapts as they grow, and eventually hands the work over~\cite{wood1976role,vandepol2010scaffolding}. Cognitive apprenticeship describes the same help from the other side: an expert makes their thinking visible, coaches the novice through it, and steps back as the novice improves~\cite{collins1989cognitive}. Our participants played both parts. We therefore argue that scaffolding between developers and AI runs both ways, and working well with AI means both learning from it and teaching it (Fig.~\ref{fig:twolearners}).

\textbf{The developer as the learner.} With AI as the more knowledgeable other, developers borrow what they lack, e.g., a skill, an expert's answer they would otherwise have had to ask for, or the capacity of a team they do not have~\cite{lee2002social}. What they gain first is reach, which is not yet understanding. A developer who stays engaged with the work can turn borrowed reach into their own skill and need less help next time~\cite{kazemitabaar2023studying,shen2026how}. A developer who does not will need the same help again, and will be no better placed to tell whether it was right.

\textbf{AI as the learner.} Conversely, developers can also act as teachers, externalizing the tacit knowledge and discernment that come with experience~\cite{polanyi1966tacit}: for example, a sense of which design will age badly, which requirement is wrong, and which fix treats a symptom rather than the cause. Since little of it is written down, AI cannot reliably learn these ways of working on its own. It reaches AI only when a developer makes it explicit, for example through a specification, a success criterion, a test, or a standing instruction such as the context files developers now write for coding agents~\cite{chatlatanagulchai2026agent,galster2026configuring}. This is the sense in which we call AI a learner. We do not imply that it has standing as a person; rather, we mean a system that performs better as humans add to the context and memory it uses~\cite {hutchins1995cognition}.

\textbf{The skill--judgment nexus.} Underlying this exchange are two things a more knowledgeable other offers: \textit{skill}, the ability to do the work, and \textit{judgment}, the ability to tell whether it is right. A human expert brings both. AI, by contrast, can supply skill beyond what it can reliably judge. We call the point where the two zones meet the \textit{skill--judgment nexus}: whatever AI produces has to be reliably judged, and developers must continuously bridge that gap.

Each time a developer supplies that judgment, the loop in Figure~\ref{fig:twolearners} can turn. For instance, picture a developer who uses AI to solve a problem slightly beyond them. If they take the time to understand the solution, they can judge the next one themselves, and their zone moves outward. If they then codify what a good solution looks like, as a test, context, or a success criterion for AI to learn from, AI can handle the next problem of that kind alone, and its zone moves outward too. The developer is now free for harder problems, which call for new judgment, and the loop turns again.

Freeing developers for harder problems raises a question the loop cannot answer on its own: which work do they want to hand over at all? Our participants handed over what they found tedious, such as the \pq{menial task that I do not enjoy working on} (P120), and held on to what they answered for: \pq{I'm the gatekeeper of my development work and I take responsibility} (P303). This matches our prior work, in which accountability and professional identity predicted which work developers kept. Developers there were least willing to let AI act on their behalf where they felt accountable for the outcome, and least willing to let it decide on their behalf in identity-defining and relationship-centric work~\cite{choudhuri2026you,choudhuri2026ai}. These are the parts of the job that people derive meaning from~\cite{rosso2010meaning}, and they are also where judgment is formed. A developer who keeps the work they answer for therefore keeps building the judgment the loop runs on. The outer ring of AI's zone in Figure~\ref{fig:twolearners} is thus as much a choice developers consciously make as a limit of current technology.

\subsection{When the loop stalls: the hollow orchestrator}\label{sec:collapse}

Keeping that work, however, takes effort, and the pace of adoption does not always leave room for it. Consider a developer whose team adopts AI quickly. Each month, AI writes more of the code and reviews more of the changes; the developer approves more and writes less. Their judgment, meanwhile, stays where it was. Our prior work names where this ends: the \textit{hollow orchestrator}, a developer who keeps the role but can no longer follow the work they approve~\cite{choudhuri2026you}.

A hollow orchestrator works on borrowed judgment, and our findings suggest they pay for it in three ways. They grow dependent: one participant called their domain knowledge \pq{the bulwark against fabrications or hallucinations impacting my work}, and worried that \pq{as my own domain knowledge gets stale, the chances that I miss something increase} (P320). Their verification burden grows, as more output arrives for review than anyone can follow~\cite{choudhuri2026copilot,mozannar2024reading}. And their expertise erodes slowly enough to go unnoticed. Ehsan et al.~\cite{ehsan2026future} followed clinicians using an AI treatment-planning system for a year and found that early efficiency gains hid ``intuition rust,'' a gradual dulling of expert judgment that later surfaced as skill atrophy and a commoditized professional identity.

The hollow orchestrator also loses meaning, and loses it through the same handover. Meaningful work is a source of enjoyment and engagement~\cite{rosso2010meaning}, and each part of the job a developer cared about and gave away takes some of that with it. Eventually, work that is harder to care about is harder to keep judging, and judgment that no one cares to exercise does not grow. At that point the developer has stopped learning, and so has stopped teaching.

When teaching stops, the loop can stall from AI's side too. Stale context misleads every run that inherits it~\cite{chatlatanagulchai2026agent}, and vague requests come back as work that looks right and is wrong~\cite{diaz2026spec}.

\subsection{Keeping the loop turning: a two-way scaffolding playbook}\label{sec:playbook}
\begin{table*}[tp]
\caption{A two-way scaffolding playbook. Each row adapts a scaffolding function from the teaching literature~\cite{wood1976role,collins1989cognitive,vandepol2010scaffolding}, grouped by when learners act, with practices for developers scaffolding AI and AI scaffolding developers, plus pitfalls to guard against.}
\label{tab:playbook}
\fontsize{7.2}{8.2}\selectfont\centering\setlength{\tabcolsep}{3pt}\renewcommand{\arraystretch}{1.0}
\newcommand{\gl}[1]{ \textit{(#1)}}
\newcommand{\pr}[1]{\textbf{#1}}
\begin{tabular}{@{}>{\raggedright\arraybackslash}p{0.125\textwidth}>{\raggedright\arraybackslash}p{0.265\textwidth}>{\raggedright\arraybackslash}p{0.14\textwidth}|>{\raggedright\arraybackslash}p{0.275\textwidth}>{\raggedright\arraybackslash}p{0.145\textwidth}@{}}
\toprule
\textbf{Scaffolding function} & \textbf{AI as the learner:} the developer scaffolds~AI & \textbf{Guard against} & \textbf{The developer as the learner:} AI scaffolds the~developer & \textbf{Guard against} \\
\midrule
\rowcolor{phBefore}[\tabcolsep][0pt]\multicolumn{5}{@{\hspace{3pt}}l@{}}{\textit{Before the learner acts} (demonstration; reduction in degrees of freedom)} \\
\addlinespace[1pt]
Demonstration~\cite{wood1976role,collins1989cognitive}\gl{showing how it is~done} & \pr{Developers show AI what good work looks like.} They pair each rule with examples of work that meets it and work that almost does, so AI learns where the line falls~($n$=4) & Exemplars that freeze yesterday's practice into every run~\cite{chatlatanagulchai2026agent} & \pr{AI walks through how an expert would approach the problem.} It lays out the key ideas behind a solution, the near misses an expert would reject, and what an expert would check~($n$=21) & Copying what one cannot yet~judge~\cite{prather2024widening} \\
\arrayrulecolor{black!22}\specialrule{0.4pt}{1.5pt}{1.5pt}\arrayrulecolor{black}
Reduction in degrees of freedom~\cite{wood1976role}\gl{simplifying the~task} & \pr{Developers size the work to what they can verify.} They split delegated work into checkable pieces, so the unit of delegation matches the unit of verification~($n$=7) & Delegating more than one can check, then approving on~trust~\cite{ehsan2026future} & \pr{AI takes the incidental work and leaves the formative.} It handles the parts of a task that do not build the skill the developer is after, and leaves them the parts that do~($n$=61) & Offloading the practice through which expertise~forms~\cite{risko2016cognitive} \\
\midrule
\rowcolor{phWhile}[\tabcolsep][0pt]\multicolumn{5}{@{\hspace{3pt}}l@{}}{\textit{While the learner acts} (marking critical features; articulation and reflection)} \\
\addlinespace[1pt]
Marking critical features~\cite{wood1976role}\gl{pointing out what~matters} & \pr{Developers keep the goal in view.} They set out the end goal, the constraints, and what must not change, in a form AI can check its own work against as it goes~($n$=5) & AI that stops before the goal is~met~\cite{lulla2026loop} & \pr{AI flags what a newcomer would miss.} It points to the risky parts, hidden assumptions, and what an expert would look at first, instead of summarizing everything evenly~($n$=33) & Missing what the summary left~out~\cite{shen2026how} \\
\arrayrulecolor{black!22}\specialrule{0.4pt}{1.5pt}{1.5pt}\arrayrulecolor{black}
Articulation and reflection~\cite{collins1989cognitive}\gl{explaining and comparing~reasoning} & \pr{Developers have AI rubber-duck its work.} AI externalizes its goal and assumptions before acting, then walks through its work, fixing issues as it goes~($n$=3) & AI judging its own output~\cite{panickssery2024llm} & \pr{AI is the developer's rubber duck.} Developers talk through their own approach with AI and then compare it with the alternatives it offers, so their reasoning and understanding improve~($n$=10) & Taking AI's comparison as~settled~\cite{lee2025impact} \\
\midrule
\rowcolor{phWrong}[\tabcolsep][0pt]\multicolumn{5}{@{\hspace{3pt}}l@{}}{\textit{When the learner goes wrong} (marking the discrepancy; frustration control)} \\
\addlinespace[1pt]
Marking the discrepancy~\cite{wood1976role}\gl{pointing out where the work is~wrong} & \pr{Developers check against an independent standard.} They judge AI's work with tests, evaluations, reviewers, or models other than the one that produced it~($n$=11) & Verifying that lapses as confidence in AI~grows~\cite{lee2025impact} & \pr{AI points to the error, not the fix.} It shows where the developer's work departs from what is expected and lets the developer correct it~($n$=4) & Letting AI find the errors one never learns to~find~\cite{shen2026how} \\
\arrayrulecolor{black!22}\specialrule{0.4pt}{1.5pt}{1.5pt}\arrayrulecolor{black}
Frustration control~\cite{wood1976role}\gl{keeping errors from derailing the~learner} & \pr{Developers make AI's failure cheap and visible.} They run AI where mistakes can be thrown away, with limits that make it stop and ask for help when stuck~($n$=3) & Runs that neither converge nor~stop~\cite{lulla2026loop} & \pr{AI makes asking safe.} Developers can ask questions they would hesitate to ask colleagues, so not knowing costs less, without lowering the standard their work must meet~($n$=8) & Comfort that turns into~dependence~\cite{wood1976role} \\
\midrule
\rowcolor{phAfter}[\tabcolsep][0pt]\multicolumn{5}{@{\hspace{3pt}}l@{}}{\textit{After the learner acts} (scaffold fading; exploration; transfer of responsibility)} \\
\addlinespace[1pt]
Scaffold fading~\cite{collins1989cognitive,vandepol2010scaffolding}\gl{adapting support as the learner~grows} & \pr{Developers extend autonomy on evidence.} They widen what AI may do alone once it has proven reliable on that kind of task, and narrow it again when it slips~($n$=7) & Autonomy granted beyond what evaluations~support~\cite{choudhuri2026you} & \pr{AI adjusts its scaffolding as the developer grows.} 
It explains less of what the developer already knows and leaves more decisions to them, adding support back as needed~($n$=7) & Support that never fades, so the skill never transfers to the~developer~\cite{risko2016cognitive} \\
\arrayrulecolor{black!22}\specialrule{0.4pt}{1.5pt}{1.5pt}\arrayrulecolor{black}
Exploration~\cite{collins1989cognitive}\gl{solving problems on one's~own} & \pr{Developers let AI explore.} When failure is cheap, AI explores freely and developers review what it found and how~($n$=35) & Unreviewed output merged into the~product~\cite{he2026speed} & \pr{Developers explore with AI.} They try their own ideas as disposable prototypes before committing to one~($n$=17) & Mistaking a working prototype for a finished~design~\cite{sarkar2025vibe} \\
\arrayrulecolor{black!22}\specialrule{0.4pt}{1.5pt}{1.5pt}\arrayrulecolor{black}
Transfer of responsibility~\cite{vandepol2010scaffolding}\gl{learner owns tasks} & \pr{Developers hand over the task and keep the decision.} AI owns routine work end to end, while a person keeps the decisions~($n$=28) & Handing over accountable or identity-defining~work~\cite{choudhuri2026you} & \pr{AI hands the specialist's work to the developer.} Developers take on work that once needed another role, and build the judgment to own it~($n$=38) & Losing the expert's~judgment~\cite{wegner1987transactive,choudhuri2026you} \\
\bottomrule
\end{tabular}
\vspace{-3mm}
\end{table*}

Keeping that teaching going raises an uncomfortable question: why would developers train systems that might replace them? We see two reasons. The first is that writing knowledge down benefits the developer and their team as much as the agent. To state what a good solution looks like, a developer has to put into words what they know but have never had to say~\cite{polanyi1966tacit}. Doing so may sharpen their own judgment, and making such knowledge explicit is how teams and organizations learn~\cite{nonaka1995knowledge}. The second is that tacit knowledge is becoming harder to pass on. Tech firms already employ fewer people~\cite{kim2026ainative}, and AI increasingly does the routine work through which juniors once built their expertise~\cite{beane2019shadow}. Knowledge that stays tacit may therefore hollow out, and the loss may surface later as various forms of debt~\cite{storey2026technical}, ranging from the erosion of a team's shared understanding of its system to the loss of the recorded goals, constraints, and rationale that guide how humans and agents change it. Externalizing tacit knowledge can reduce both, because the specifications and context that teach the next agent can also teach the next developer. Seen this way, codified judgment is less a gift to the machine than a record of the developer's own expertise. Should they be able to carry it with them, the way a professor who moves to a new university takes their course materials along? Who owns such records is still an open question~\cite{lindley2023building}, and organizations that want developers to externalize their tacit knowledge will need to answer it. They will also need to make time for this teaching. Just as electrification paid off only once factories were rebuilt around it~\cite{david1990dynamo}, the loop will pay off only if teams make that time.

Based on our findings and the scaffolding literature, Table~\ref{tab:playbook} sets out a two-way playbook for keeping the loop turning: nine functions from research on tutoring~\cite{wood1976role}, cognitive apprenticeship~\cite{collins1989cognitive}, and fading support~\cite{vandepol2010scaffolding}, grouped by when the learner acts: with a practice for developers teaching AI, a practice for AI teaching developers, and common pitfalls to guard against. Three practices merit particular discussion. 

The first is \textit{demonstration}. Developers can convey their work standards to AI through contextualized examples. For instance, an underspecified instruction such as ``handle errors better'' conveys little, whereas a pair of examples, one that conforms to the team's conventions and one that nearly does, helps AI infer what the team actually expects. Similarly, AI can demonstrate to a developer who is new to a problem how an expert would approach it, including the key ideas behind a solution, the near misses an expert would reject, and what an expert would actually check. 

The second is \textit{articulation}, drawing upon rubber-duck debugging. Here, developers can articulate their approach by thinking aloud with AI before they commit to a solution. In turn, AI can rubber-duck its own work: before acting, it states the goal and the assumptions it relies on, and after acting, it walks through what it built. 

The final one concerns \textit{what lasts after a task}, which is what turns the loop from help in the moment into durable learning. Exploration shows one way how. Where failure is cheap, developers can give AI a goal rather than steps and let it find its own route; the exploration becomes durable when the developer folds what worked, and what did not, back into the context and success criteria that the next work inherits. Developers explore the same way, trying their own ideas as prototypes before committing to one, and each discarded prototype leaves them with judgment they didn't have before. 


The specific practices are likely to change as models improve and developers' experience evolves, and the same function can take different forms for different tasks, teams, and learners. The functions underneath them, however, are likely to outlast any particular generation of tools. What they give teams and tool builders is a set of questions to ask: who is scaffolding whom, how much support does each learner need, and how should that support change as both learners grow?

\subsection{Implications for future research}\label{sec:research}

Our findings raise questions for the future of software work in the AI era. The first concerns how the nexus itself changes: does it move as models improve, or does each gain in skill open a new gap in judgment?

The second concerns whether a stalled loop can restart. Can a hollow orchestrator rebuild their judgment, and what would it take? Because expertise erodes slowly, research could look for early signs of that drift and test whether practices like those in Table~\ref{tab:playbook} slow or reverse it.

The third concerns how developers become experts. If AI now does much of the work through which expertise has always developed, how will the next generation of developers build it, and what should software engineering education teach them? Answering this also calls for rethinking AI literacy as a two-way street: knowing how to learn from AI, where to use it, how to teach it, and what work to keep.
Curricula may need to teach students effective ways of judging what AI builds and to explain what they know to an agent that will act on it.

\section{Conclusion}
\label{sec:conclusion}

Our study uncovered the different ways AI helps developers in their work: the many avenues through which it makes familiar work faster, the work it makes possible that once needed skills developers lacked or time they could not justify, alongside the new work it creates around directing, supervising, and building for AI agents. Viewed through the zone of proximal development, these changes cast developers and AI as joint learners who meet at the skill--judgment nexus, where AI supplies skill faster than it can reliably judge it and developers must continuously close that gap with judgment.

The future of AI-native software development depends on whether developers can keep closing that gap, growing their judgment as they learn with AI and externalizing their knowledge so that AI, and the developers who follow them, can learn from it. Like the riders on that train in 1865, we cannot yet see the whole route. How far we go will depend as much on the next frontier model as on what developers keep learning and building, and on what they choose to pass on.

\bibliographystyle{IEEEtran}
\footnotesize{\bibliography{references}}

\end{document}